\documentclass[preprint, aps,graphicx,prb, showpacs]{revtex4}
\usepackage{amssymb}
\usepackage{amsfonts}
\usepackage{amsmath}

\usepackage{graphicx}

\begin{document}

\title{Wavelength-Scale Imaging of Trapped Ions using a Phase Fresnel lens}

\author{A. Jechow,$^*$ E. W. Streed, B. G. Norton, M. J. Petrasiunas, D. Kielpinski}

\affiliation{Centre for Quantum Dynamics, Griffith University, Nathan, QLD 4111, Australia \\
$^*$Corresponding author: andreas.jechow@gmx.de
}

\begin{abstract}A microfabricated phase Fresnel lens was used to image ytterbium ions trapped in a radio frequency Paul trap. The ions were laser cooled close to the Doppler limit on the 369.5 nm transition, reducing the ion motion so that each ion formed a near point source. By detecting the ion fluorescence on the same transition, near diffraction limited imaging with spot sizes of below 440 nm (FWHM) was achieved. This is the first demonstration of imaging trapped ions with a resolution on the order of the transition wavelength.\end{abstract}

\date{\today}


\maketitle

Laser cooled trapped atomic ions are a nearly ideal system to test fundamental atomic and quantum physics at very high precision. They are effectively isolated atoms held at rest and largely free from perturbations, representing a quantum system with control over all degrees of freedom \cite{Leibfried03QD_rev}. Thus, they have been used for studies in many fields of classical and nonclassical physics most prominently in quantum computing \cite{cirac-95,Kielpinski-02}, quantum information processing (QIP) \cite{garcia-ripoll-05,Amini-10,Kielpinski-03} and precision metrology\cite{chou10, biercuk}. Very recently, trapped ions have been used to simulate the Dirac equation \cite{rooszitter} and to demonstrate the phonon laser\cite{udemphononlaser}.

While most of the technology for laser cooling and trapping has been established for decades\cite{ionspaultrap}, it is still challenging to obtain high resolution images of trapped ions. Due to the small size and the highly diverging emission pattern of the ion, an imaging system with low aberrations and a large numerical aperture (NA) is required. Typically, multi-element objectives are used for this purpose, achieving resolutions on the order of 1$\, \mu$m \cite{Leibfried03QD_rev}. However, increasing the NA of these lens systems results in very short working distances leading to charging effects \cite{harlander}. Recent attempts to overcome this issue have used bulk parabolic and spherical mirrors \cite{Maiwald-09,Sondermann-07,Shu-10}, but high resolution imaging has not been demonstrated yet.

As part of a roadmap towards large scale QIP with trapped ions \cite{Amini-10, Kielpinski-02, Kielpinski-03}, we have proposed the use of phase Fresnel lens (PFL) arrays as a scalable optical interconnect \cite{Streed-09}. PFLs are diffractive optics that can provide both, a high NA and diffraction limited imaging. Conventional optics cannot be scaled to large arrays and are bulky and cost intensive. In contrast, PFLs can be microfabricated in large quantities at low costs and have been widely used in other areas of optics e.g. for the collimation and beam quality improvement of diode laser arrays \cite{leger}.

Recently, we successfully demonstrated the first proof of principle experiment on PFL imaging ions by integrating a single microfabricated PFL with a radio frequency (RF) ion trap in an ultra-high vacuum (UHV) chamber \cite{erik}. In that experiment, the image resolution was limited to about 4$\, \mu$m by residual ion motion.

Here we present recent data with an improved setup. Ion images with wavelength-scale resolution were obtained for the first time, demonstrating near-diffraction-limited performance of the PFL. This high resolution enables the possibility of individual addressing and individual readout of the ions by lasers, as required for large-scale ion-trap QIP \cite{naegerl-99,duan-10}. While our approach mainly targets large-scale QIP with trapped ions, it can be easily transferred to neutral atoms and solid state QIP applications \cite{hadden-10}.
\begin{figure}[t]
\centerline{\includegraphics[width=0.6\columnwidth]{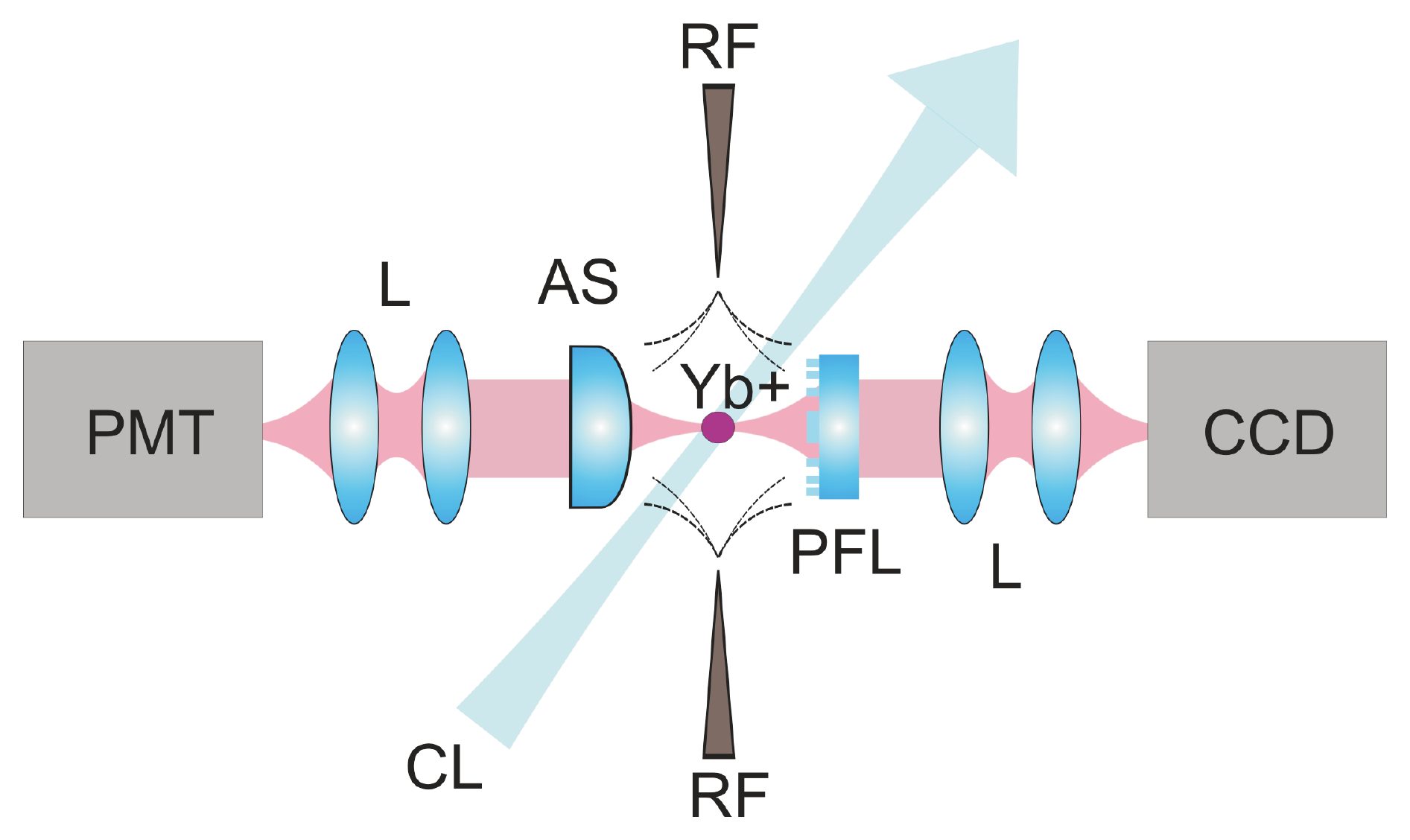}}
\caption{Schematic of the experimental setup. The 369.5$\,$nm light scattered by a trapped Yb ion is collected by the PFL for imaging on a CCD camera. The PMT is used to detect the ion fluorescence rate. (PFL - phase Fresnel lens, AS - aspheric lens, RF - tungsten needles with RF field applied, CL - cooling laser beam, L - imaging lens systems, PMT - photo multiplier tube).}
\end{figure}

The experimental setup is schematically depicted in Fig. 1. A PFL was integrated with a needle trap to collimate the fluorescence of trapped Yb$^+$ at 369.5$\,$nm and image it onto a cooled charge coupled device (CCD) camera. An aspheric lens was used to detect the ion fluorescence rate with a photo multiplier tube (PMT).

A two-needle Paul trap design \cite{Deslauries-06} was used to form a RF electric quadrupole field. A potential $V_0 \cos(\Omega_{RF}t)$ with $V_0 \approx 200$ V, $\Omega_{RF}/2\pi = 20$ MHz was applied between two tungsten needles having a tip diameter of about 10$\, \mu$m and a spacing of approximately 300$\, \mu$m. Four additional needles are located close to the trapping region to allow for the compensation of stray electric DC fields. The background pressure in the vacuum chamber was measured to be $3.7\times10^{-11}$ mbar.

We used nano-positioning translation stages to precisely control the positioning of the ion with respect to the PFL. The needles were attached to flexible mechanical feedthroughs by ceramic insulators to guarantee independent movement. The stages were located outside the vacuum chamber and attached to piezo motor actuators and dial indicators. This allowed positioning in all three dimensions with a precision of better than 100$\,$nm.
 
$^{174}$Yb$^+$ ions were loaded into the trap by isotope-selective photoionization of a neutral ytterbium atomic beam \cite{Streed-08}. Laser cooling was performed on the 369.5$\,$nm S$_{1/2}$ to P$_{1/2}$ transition using light from an external cavity diode laser (ECDL) \cite{Kielpinski-06}. The laser was frequency stabilized with a dichroic atomic vapor laser lock (DAVLL) to $\mbox{Yb}^+$ ions generated in an electrical discharge \cite{Streed-08}. The cooling laser was focused to a $1/e^2$ beam waist diameter of 80$\, \mu$m. The ions were repumped from the metastable F$_{7/2}$ and D$_{3/2}$ dark states using ECDLs at 638$\,$nm and 935.2$\,$nm, respectively. With this system ions with lifetimes of several hours were observed.

The PFL had a focal length of $f$=$\,$3$\,$mm, the same as its working distance, and had a clear aperture of $d$$\,$=$\,$5$\,$mm diameter. The lens therefore had a speed of $f$/0.6 and a NA$\,$=sin$\theta$\footnote[1]{in contrast to the often used approximation NA$\approx d/2f$ which would give an NA of 0.83 in our case}=$\,$0.64, which corresponds to 12\% of the total solid angle. The PFL was designed as a binary grating structure with a step profile. It was fabricated by electron-beam lithography on a fused silica substrate at the Fraunhofer-Institut f\"ur Nachrichtentechnik, Germany. The e-beam patterning was used to write a series of concentric rings, according to the scalar design equation for a binary PFL. The rings were etched to a depth of 390$\,$nm, corresponding to contours with a $\pi$ phase step. The PFL was characterized with a knife edge beam profiling method to show diffraction limited performance \cite{Chapman-07}. More details about the lens and the fabrication can be found in \cite{Chapman-07, Streed-09, erik}.

The RF needles were positioned such that the PFL collimated the light scattered from the trapped ions. The ion fluorescence was collected by the PFL and imaged onto the camera (Andor Model DV437-BU2) with a magnification of 615$\,\pm\,$9. The camera has an array of 512x512 pixels with a pixel size of 13$\, \mu$m$\,$x$\,$13$\, \mu$m. It was cooled to $-30\,^{\circ}{\rm C}$ to reduce readout noise.
\begin{figure}[h]
\centerline{\includegraphics[width=4.4cm]{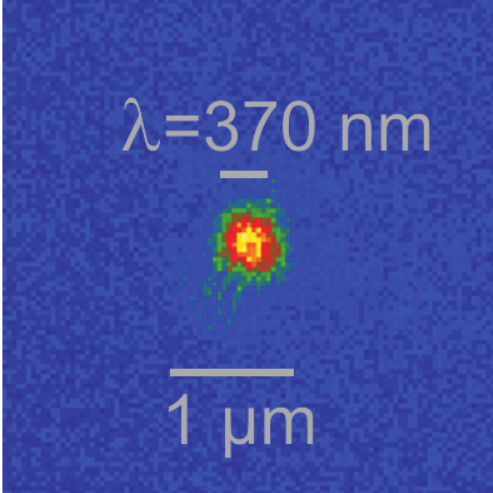} \includegraphics[width=1.7cm]{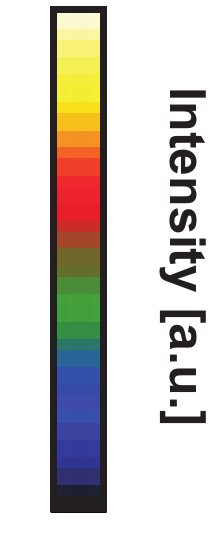}}
\caption{Image of a single ion obtained with a PFL integrated in UHV of a needle ion trap. A Gaussian fit to the data gives an ion spot size (FWHM) of 434$\,$nm$\, \pm\,$9$\,$nm in the vertical axis and 475$\,$nm$\, \pm\,$9$\,$nm in the horizontal (needle) axis.}
\end{figure}

An image of a single ion on the CCD camera is shown in Fig. 2, where the horizontal axis is the needle axis. We measured the lineshape of the 369.5$\,$nm transition by tuning the cooling laser and detecting the fluorescence counts with the PMT. The lineshape indicated a temperature close to the Doppler limit and we infer a residual ion motion amplitude of less than 15$\,$nm RMS transverse to the needle axis. Hence the ion acted as a near-ideal point source.

Magnification calibration was performed by measuring the distance between two ions for a known trap frequency. To measure the ion spacing we applied a DC voltage to the RF needles so that the ions lay in the object plane of the imaging system. In this case, the trapping potential was strong perpendicular to the needle axis and weak along the needle axis. 

Fig. 3 shows an image of two ions aligned along the needle axis. The ion spacing $l$ due to Coulomb repulsion is given by \cite{james}:
\begin{equation}
l = {\left( {{e^2 \over {8 \pi^3 \epsilon_0 M \nu^2}}} \right)^{1/3} },
\end{equation}
with the mass of each ion $M$, the electron charge $e$, the permittivity of
free space $\epsilon_0$ and the trap frequency along the needle axis $\nu$.

\begin{figure}[h]
\centerline{\includegraphics[width=6.8cm]{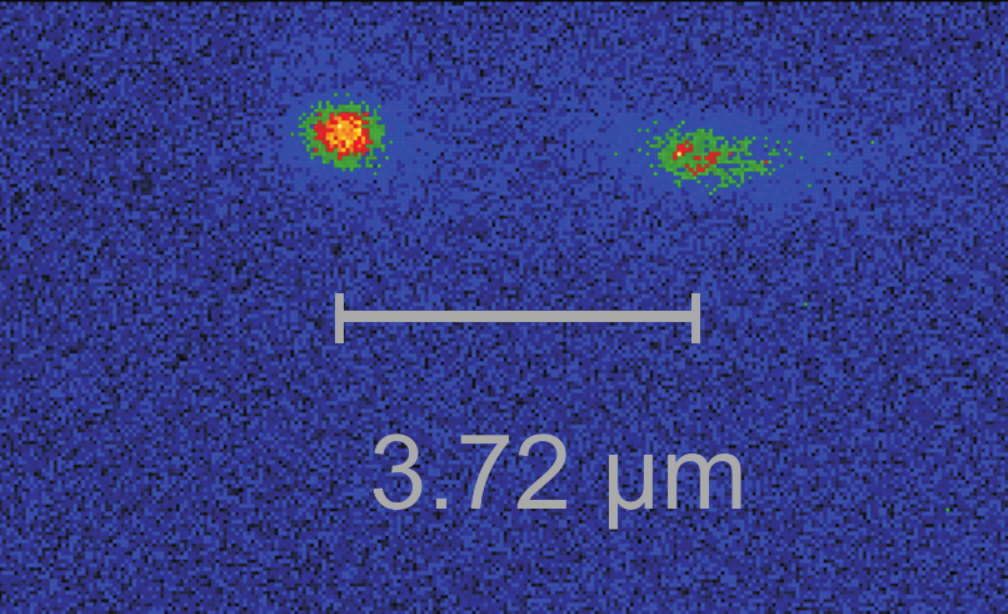} \includegraphics[width=1.6cm]{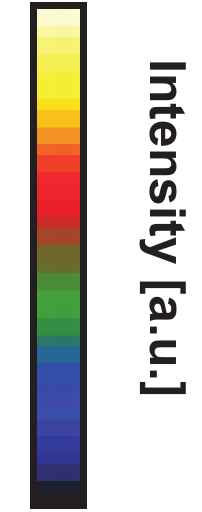}}
\caption{Image of two ions aligned in the needle axis (horizontal axis). A trap frequency $\nu\,$=882$\,\pm\,$2 kHz was measured and an ion to ion distance of 3.72$\,\pm\,$0.01$\, \mu$m was calculated from equation (1) and used for magnification calibration.}
\end{figure}
The trap frequency was determined by resonantly exciting the ion motion by applying an AC voltage to one of the compensation needles and observing a change in the fluorescence rate.  A trap frequency of $\nu\,$=882$\,\pm\,$2 kHz was measured. For $^{174}$Yb$^+$, this corresponds to an ion-ion spacing of 3.72$\,\pm\,$0.01$\, \mu$m. The resulting magnification was calculated to be 615$\,\pm\,$9. To validate the magnification calibration the ion was displaced by moving the needles. The displacement was measured with the dial indicators and compared with the displacement of the ion image on the CCD camera. The two techniques showed an agreement of better than 5$\%$.

The size of the ion image (FWHM) in Fig. 2 was calculated using Gaussian fits through the data. The ion extent was 434$\,$nm$\, \pm\,$9$\,$nm in the vertical axis and 475$\,$nm$\, \pm\,$9$\,$nm in the horizontal axis. This represents an improvement of an order of magnitude over our previous results \cite{erik}. In our setup, laser cooling was less efficient along the needle axis than along other directions. This can lead to an increase of residual ion motion and a slight blurring of the ion image along the needle axis. This effect is investigated in more detail in forthcoming work \cite{norton-10}.  

In conclusion, we have demonstrated imaging of trapped ions with a resolution at the wavelength scale. To our knowledge, this is the highest imaging resolution achieved with an atom in free space to date. The highest resolution previously achieved with imaging of single atoms was reported to be 570$\,$nm \cite{bakr} at a wavelength of 780$\,$nm. However, in that setup neutral atoms in an optical lattice have been investigated. While our approach can be applied to such neutral atomic systems fairly straightforwardly, the working distance in \cite{bakr} was too short to be applicable to trapped ions. The excellent scalability and the high resolution of the PFL architecture render it useful for integrated imaging systems in large-scale QIP applications.

This work is funded by the Australian Research Council under DP0773354 (DK), DP0877936 (ES, Australian Postdoctoral Fellowship), and FF0458313 (H. Wiseman, Federation Fellowship), as well as the US Air Force Office of Scientific Research (FA2386-09-1-4015). AJ is supported by a Griffith University Postdoctoral Fellowship. The PFL was fabricated by Margit Ferstl of the Heinrich-Hertz-Institut of the Fraunhofer-Institut f\"{u}r Nachrichtentechnik in Germany.


\end{document}